\documentclass[submission,copyright,creativecommons]{eptcs}
\providecommand{\event}{QSQW 2020}
\usepackage{underscore}
\usepackage{graphicx}
\usepackage{amsmath,amssymb}
\usepackage{cite}
\newcommand{\ket}[1]{|{#1}\rangle}

\makeatother
\makeatletter
\title{Detecting Temporal Correlation \\ via Quantum Random Number Generation}
\author{
Yutaka Shikano
\institute{Quantum Computing Center \\ Keio University \\ Yokohama, Japan}
\institute{Institute for Quantum Studies \\ Chapman University \\
California, USA}
\email{yutaka.shikano@keio.jp}
\and
Kentaro Tamura
\institute{Department of Applied Physics and Physico-Informatics \\
Keio University \\ Yokohama, Japan}
\email{cicero@keio.jp}
\and
Rudy Raymond
\institute{IBM Research -- Tokyo \\ Tokyo, Japan}
\institute{Quantum Computing Center \\ Keio University \\ Yokohama, Japan}
\email{rudyhar@jp.ibm.com}
}
\def\titlerunning{Detecting Temporal Correlation via Quantum Random Number Generator}
\def\authorrunning{Y. Shikano, K. Tamura \& R. Raymond}

\begin{document}
\maketitle

\begin{abstract}
All computing devices, including quantum computers, must exhibit that for a given input, an output is produced in accordance with the program. The outputs generated by quantum computers that fulfill these requirements are not temporally correlated, however. In a quantum-computing device comprising solid-state qubits such as superconducting qubits, any operation to rest the qubits to their initial state faces a practical problem. We applied a statistical analysis to a collection of random numbers output from a 20-qubit superconducting-qubit cloud quantum computer using the simplest random number generation scheme. The analysis indicates temporal correlation in the output of some sequences obtained from the 20 qubits. This temporal correlation may be not related to the relaxation time of each qubit. 
\end{abstract}

\section{Introduction}
The unpredictable output of random number generators (RNGs) is essential for secure communications and is a key cryptographic assumption when proving the security of encrypted communications from an information-theoretical viewpoint. While several statistical tests like the NIST Test Suites~\cite{NIST} and TestU01~\cite{testu01} are used in practice, passing these tests is a necessary but not sufficient condition to confirm the unpredictability of an RNG. For a digital programmable computer, the outcome of a computation is deterministically computed from the code and the input. While many randomized algorithms have been used to compensate for the deterministic nature of digital programmable computers, these algorithms can only ever yield pseudo-RNGs, which means that although their outputs are hard to distinguish from true RNGs under some computational-hardness assumption, they are theoretically computable from the inputs. 
Pseudo-RNGs are functions that produce an output number given an input, the latter of which is called the {\it seed}. Pseudo-RNGs are ultimately predictable because of this seed. 
Given that pseudo-RNGs are predictable, physical RNGs have been proposed and implemented for commercial use. However, physical RNGs are also theoretically predictable when they are based on classical physics, because all particles dynamics are predictable if all input parameters are known. In macroscopically described  theories such as thermodynamics or statistical physics, statistical descriptions are used to simplify the treatment of a huge number of particles. Even with unlimited computational capability, any macroscopic description must be consistent with the microscopic dynamics of all particles involved. Therefore, classical physical phenomena are, in principle, predictable. 
On the other hand, quantum random number generators (QRNGs) may overcome the problem of predictability because the measurement outcomes in quantum mechanics are associated with the quantum state only in terms of probabilities. This is called the Born rule, and serves as one of the mathematical axioms of quantum mechanics. For the practical application of RNGs, rapid and reliable operation is required. QRNGs are often implemented in quantum optics, as seen in recent review papers~\cite{Ma2016,Herrero-Collantes2017}. 
Currently available quantum computing devices consist of a small number of fully controllable integrated qubits. A quantum computer can therefore serve as an unbiased QRNG. The gate operation and measurements in such a computer may be deterministic, but the outcome will be probabilistic. Unbiased output sequences can be obtained from an equal-weight superposition of all qubits in the computer. This kind of computer does not require any input randomness, and can therefore be regarded as a \textit{seedless} RNG. A QRNG is an application of a quantum computer that requires only one qubit. This computer cannot be implemented with conventional digital computers because conventional digital computers are fundamentally incapable of performing truly random number generation.

To acquire long random output sequences, a quantum circuit needs to be repeated many times. Because the measurement changes the quantum state of a qubit, a reset operation is required after every measurement to recover the initial state. This challenge also arises when running quantum algorithms on a quantum computer, as almost all algorithms require the same quantum circuit to be run multiple times~\cite{Kak1999}. The initial state should be reset to for any runs of algorithms. It is necessary that none of the output sequences is temporally correlated. According to Landauer's principle, the memory reset operation, in principle, has an energy cost~\cite{Landauer}. This principle only holds, however, in case that computation has no energy cost~\cite{Hosoya2011}. The preceding outcome can be predicted when the additional energy cost required for the reset operation is associated with the state. When the initial state has each qubit in its ground state, the simplest reset operation is simply waiting for the qubits to return to their initial states. In solid-state qubits such as a superconducting qubit, the relaxation time ($T_1$) is typically used as the timescale for the reset operation. To let the system relax near to the ground state, the waiting time is set to more than $10$ times the value of $T_1$. Also, the relationship between relaxation and the coherence time has been discussed~\cite{Petta2005}. Fast and efficient protocols for the reset operation have been demonstrated recently~\cite{Riste2012,Tuorila2017,Egger2018,Magnard2018,basilewitsch2020}. 

This paper aims to uncover the relationship between the temporal correlation of output random number sequences and the relaxation time of each qubit.
Another possible source of temporal correlation in the output sequences is hardware-correlated noise. Regarding hardware system identification, a comprehensive review~\cite{Eisert2019} lists several methodologies. Methods that use quantum random number generation and statistical random-number analysis cannot be used to identify the source of systematic noise.

\section{Quantum Random Number Generation in Quantum Computers}
The simplest procedure for quantum random number generation in a quantum computer is explained below. First, we prepare the initial state $\ket{0}$. Second, we apply a Hadamard gate to the initial state, creating the superposed state $\frac{1}{\sqrt{2}}(\ket{0} + \ket{1})$. Third, we measure this superposed state in the computational basis $\ket{0}$ and $\ket{1}$. According to the mathematical axiom of quantum mechanics, the measurement outcome should be uniformly random. Otherwise, the probabilistic structure of quantum mechanics would be unnecessary. Because the randomness-generation process relies on the fundamental axiom of quantum mechanics, quantum random number generation is controlled in the same way every time. Therefore, such random number generator is \textit{seedless}. This procedure is repeated to yield a random-number sequence. When each step of the procedure is independent, the generated output sequence is independent identically distributed (i.i.d.). For a quantum computing device with several qubits, each qubit can generate output sequences in parallel with the same quantum circuit. 

Several implementations of QRNGs on programmable quantum computers have been tested~\cite{tamura2019, tamura2020}. Since none of the generated output sequences are ideal random numbers without applying information processing, programmable quantum computers available today give noisy results. Therefore, the output sequences of the quantum-circuit based QRNGs include information about the quantum-computing device, such as the stability of the operation~\cite{tamura2020}.

\section{Autocorrelation Test} \label{sec:ind}
An ideal RNG is equivalent to a random variable $\{X_t\} \ (t = 1, 2, \cdots, n)$ that produces independent and uniform bits. 
TestU01 is a collection of empirical statistical tests for RNGs that indicate whether an RNG is ideal~\cite{testu01}. The autocorrelation test in TestU01 checks whether pairs of bits that are $l$ bits apart within a bit sequence are independent of each other. The test statistic for an obtained binary sequence $x_1, x_2, \cdots x_n$ is defined as
\begin{align}
    A_l(x_1, x_2, \cdots x_n) \equiv \sum_{i = 1}^{n - l} x_i\oplus x_{i + l}.
    \label{eq:autocorr}
\end{align}
By setting $l = 1$, the autocorrelation test becomes a test for independence between neighboring bits. Given that the test statistic $A_l$ follows the binomial distribution where the number of trials is $n-l$ and the probability of obtaining the output ``1" is $\tilde{p}$, $A_l$ is approximately normal when $n-l$ is large~\cite{testu01}. The test statistic $A_l$ is converted to $A_l'$ so that it follows the standard normal distribution as
\begin{align}
    A_l' \equiv \frac{A_l - 2\tilde{p}(1-\tilde{p})(n - l)}{\sqrt{2\tilde{p}(1-\tilde{p})(n-l)(1-2\tilde{p}(1-\tilde{p}))}}. \label{new}
\end{align}

Under the assumption of identical random variables $\{ X_t \} \ (t = 1, 2, \cdots, n)$ with the probability $\tilde{p}$ obtaining the output ``1", the null hypothesis $H_0$ is set as 
\begin{quote}
    $H_0$: $\{ X_t \} \ (t = 1, 2, \cdots, n)$ is independent, that is, not correlated.
\end{quote}
Since $A_l'$ approximately follows the standard normal distribution, the one-sample, two-sided z-test is applied to obtain the p-value for the null hypothesis $H_0$ as 
\begin{equation}
    \operatorname{p-value} = \mathrm{erfc}\left(\frac{A_l'}{\sqrt{2}}\right),
\end{equation}
where the complementary error function is defined as
\begin{align}
    \mathrm{erfc}(x) \equiv \frac{2}{\sqrt{\pi}}\int_x^{\infty}e^{-t^2}dt.
\end{align}
When the p-value is less than $\alpha = 0.01$, all $l$-separated bits within the sequence are regarded as correlated. Otherwise, the sequence is regarded as independent. $\alpha$ is called the level of significance, and it is the minimum p-value of a sample that we accept as likely to have been produced by an ideal RNG. Here, the level of significance is set at $\alpha = 0.01$, which is the standard setting among statistical tests for RNGs. The failure of this test $H_0$ indicates that the biased sequence $\{ X_t \} \ (t = 1, 2, \cdots, n)$ has temporal correlation.
As one expects an ideal RNG to produce a sample with a p-value under the level of significance with probability $\alpha$, the proportion of p-values equal to or greater than $\alpha$ provides a more comprehensive indicator of the behavior of an RNG than a single p-value does.

\begin{figure}[bt]
    \centering
    \includegraphics[width=16cm]{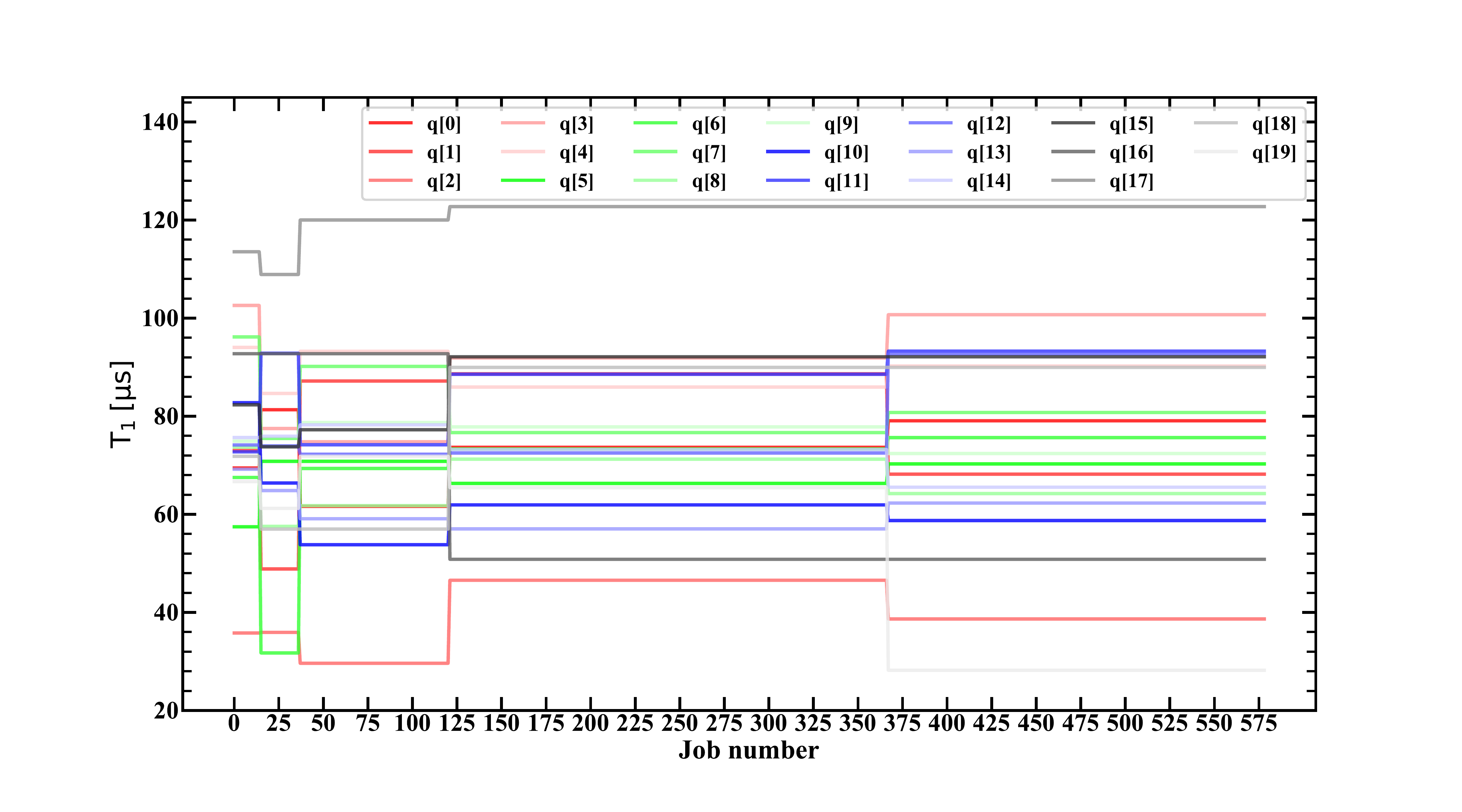}
    \caption{$T_1$ time of each qubit provided by Qiskit~\cite{qiskit}.}
    \label{fig:t1}
\end{figure}
\section{Temporal Correlation Detection in Quantum Computer}
As a state-of-the-art experiment, we applied this QRNG procedure to the 20-qubit superconducting qubit device called the IBM Q Poughkeepsie which is connected via the cloud service~\cite{tamura2020}. The sequences analyzed in this paper were the same as those in Ref.~\cite{tamura2020}. The data were taken from the device by generating 8192 bits per job and repeating the process. All jobs were sent during the time span from 2019/05/09 11:24:27 (GMT) to 2019/05/12 23:24:58 (GMT). The calibration time can be seen in Ref.~\cite[Table 6]{tamura2020}. 579 jobs were run over the course of five days. The $T_1$ time of each qubit was measured and provided as device information using an open-source framework for working with noisy quantum computers at the level of pulses, circuits, and algorithms, known as Qiskit~\cite{qiskit}. This analysis tool is illustrated in Fig.~\ref{fig:t1}\footnote{We cannot directly evaluate the reset time due to system regulation and cannot guess whether the $T_1$ of a qubit affects temporal correlation. According to private communications with IBM Q Network Support team, each single circuit is operated at the usual $1$ kHz repetition rate, which means 1/repetition rate = $1$ ms for one circuit execution. On IBM Q Poughkeepsie device, a single circuit execution consists of one initialization step, the quantum gates, the measurement and the relaxation time (dead time before reaching 1/repetition rate). Moreover, four calibration circuits are executed between each circuit execution which takes around $4$ ms.}.
In reality, $T_1$ fluctuates, according to a report about a different device~\cite{Burnett2019}.
\begin{figure}[t]
    \centering
    \includegraphics[width=15cm]{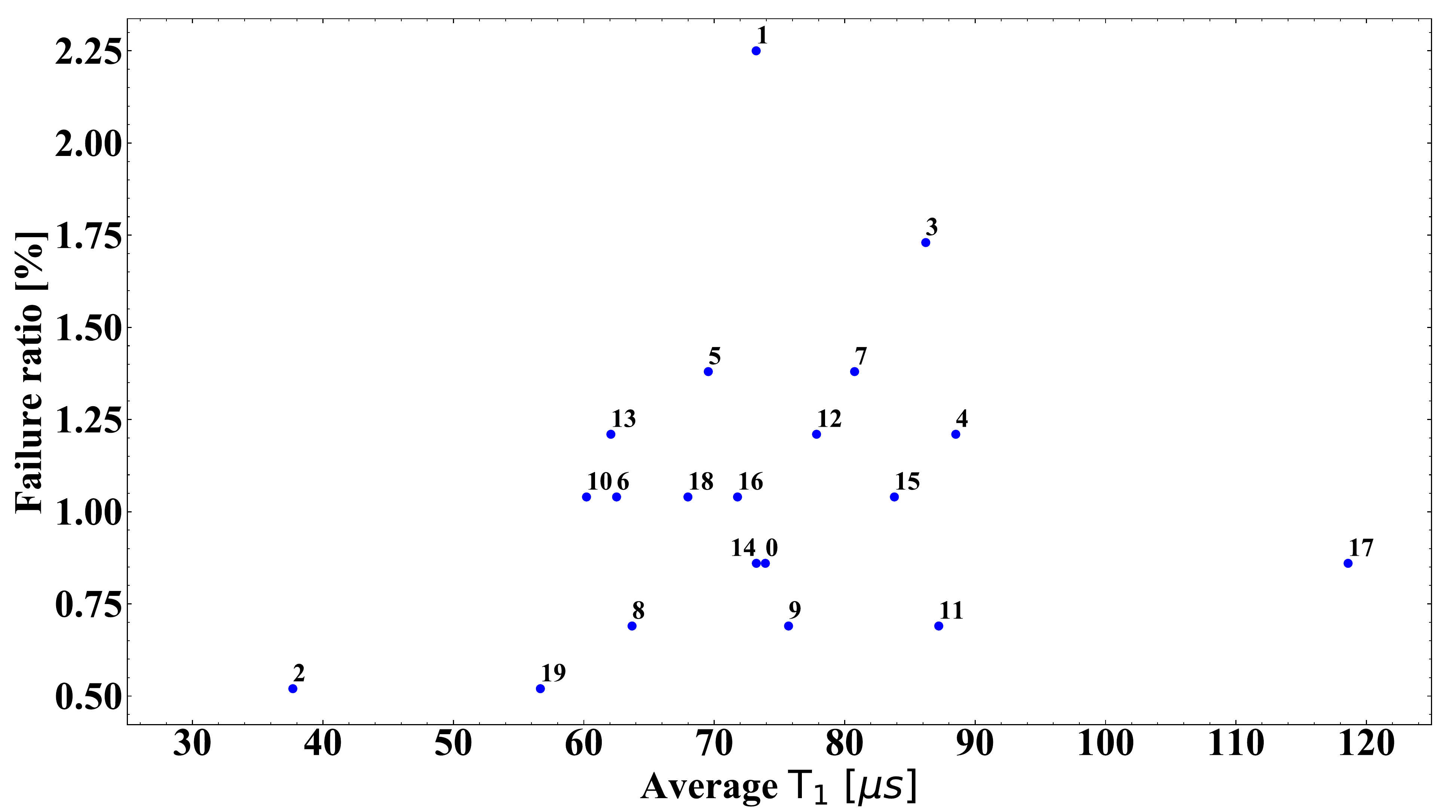}
    \caption{Failure ratio of the autocorrelation test for each qubit along with the average $T_1$. The label of the point represents the qubit number. The time change of the $T_1$ time is seen in Fig.~\ref{fig:t1}.}
    \label{fig:auto}
\end{figure}
\begin{figure}[p]
    \centering
    \includegraphics[width=15.7cm]{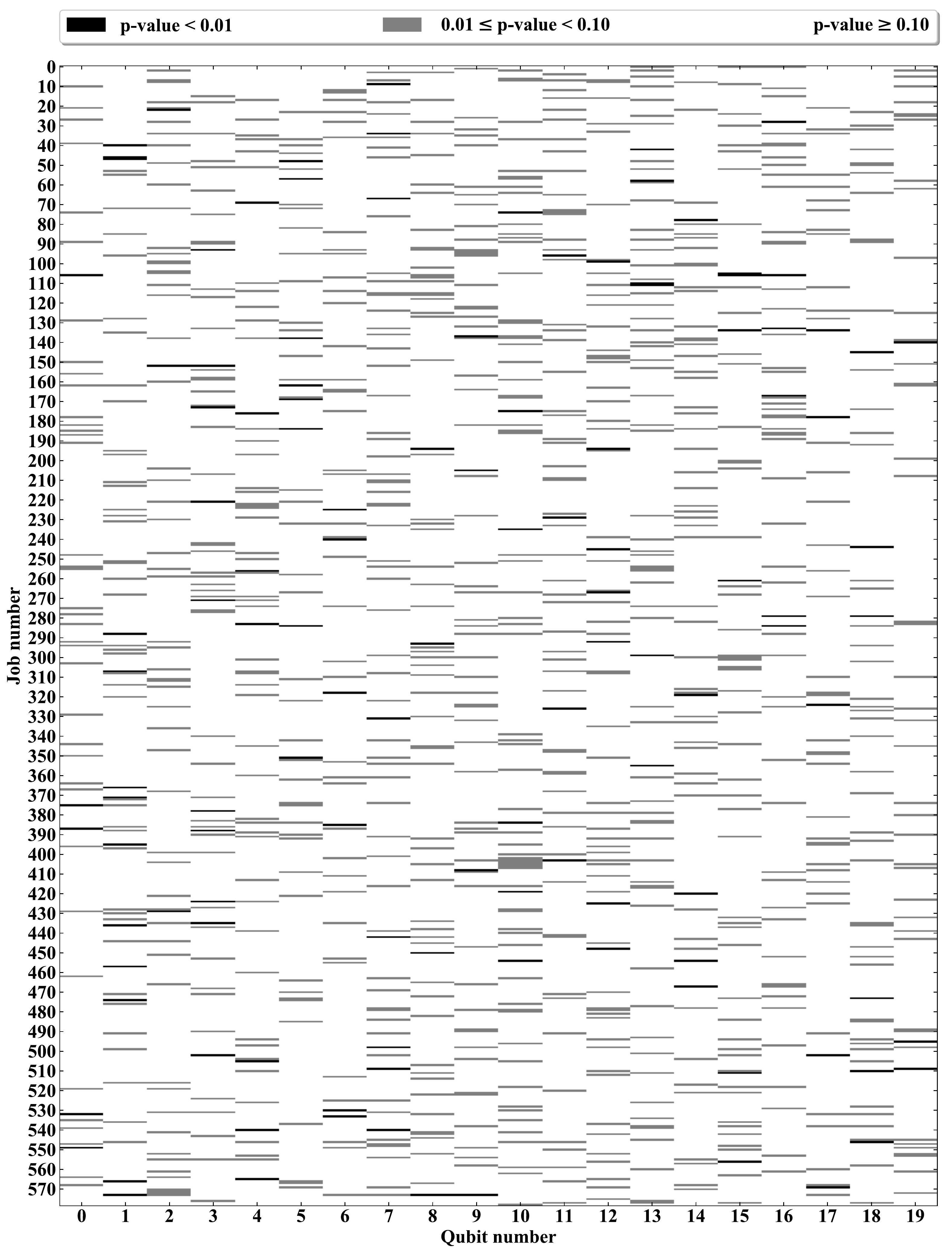}
    \caption{p-values of the autocorrelation test for each qubit (x-axis) at each job performed (y-axis) .}
    \label{fig:pvalue}
\end{figure}

Let us apply the autocorrelation test in TestU01 as explained in Sec.~\ref{sec:ind} to the output sequence generated by each qubit. The value of $\tilde{p}$ at Eq.~(\ref{new})\footnote{It is noted that for ideal quantum device, the value of $\tilde{p}$ at Eq.~(\ref{new}) is the unbiased $1/2$ as aforementioned before, but we allow the value of $\tilde{p}$ to be biased considering the noisy quantum devices.} is estimated to be a frequency probability for each job and each qubit\footnote{The ideal distribution per a qubit is assumed for the samples within the job. This means that a quantum device to obtain the generated 8192 samples within the same job per a qubit is stably operated and is not changed on the hardware information within the same job. The min-entropy calculated from the value of $\tilde{p}$ is seen in Ref.~\cite[Fig. 2]{tamura2020}.}. Figure~\ref{fig:auto} displays the number of failures under the $1$\% confidence level of the autocorrelation test in the case of $l=1$ along with the average $T_1$ for each qubit and shows no apparent relationship between the failure ratio of the autocorrelation test and $T_1$. Therefore, the temporal correlation of the output sequences does not come from an imperfection in the reset operation. This result suggests the existence of hardware imperfections or systematic error in the quantum computing device. Also, $81.0$\% of the jobs pass the autocorrelation test for qubits simultaneously. The summary of all p-values of the autocorrelation test for each qubit are listed in Fig.~\ref{fig:pvalue}. 

\section{Conclusion} \label{sec:conclusion}
A quantum random number generator is an essential component for ultimate secure encryption. Since a QRNG is a seedless RNG, it is crucial that the costs of repeated operations be independent of the outcome. As a one-qubit application of a quantum computer, we connected to a superconducting quantum computer via the cloud to execute the simplest quantum random number generation scheme. Statistical analysis of the results showed that the random number sequences output by the computer were biased and that some sequences had temporal correlation. This temporal correlation, combined with the instability of the quantum computer~\cite{tamura2020}, shows that the quantum random number generation by the quantum computing device is far from the ideal one. Further research on how much quantum algorithms are affected by temporal correlations is needed. These correlations should be eliminated or negligible small for large-scale quantum computation.

\section*{Acknowledgements}
The authors thank Hermanni Heimonen, Gregor Weihs, Atsushi Iwasaki, and Hidetoshi Okutomi for valuable discussion. Some of the authors (Y.S., K.T.) are also grateful to Patrick Mensac and Francois Varchon of the IBM Q Network support team for sharing information on the qubit initialization in IBM Q systems. This work is partially supported by JSPS KAKENHI (Grant Nos. 17K05082 and 19H05156) and JST, PRESTO (feasibility study of specific research proposal) Grant Number JPMJPR19MB. The results presented in this paper were obtained in part using an IBM Q quantum computing system as part of the IBM Q Network. 

\nocite{*}
\bibliographystyle{eptcs}
\bibliography{autocorrelation}

\begin{thebibliography}{10}
\providecommand{\bibitemdeclare}[2]{}
\providecommand{\surnamestart}{}
\providecommand{\surnameend}{}
\providecommand{\urlprefix}{Available at }
\providecommand{\url}[1]{\texttt{#1}}
\providecommand{\href}[2]{\texttt{#2}}
\providecommand{\urlalt}[2]{\href{#1}{#2}}
\providecommand{\doi}[1]{doi:\urlalt{http://dx.doi.org/#1}{#1}}
\providecommand{\bibinfo}[2]{#2}

\bibitemdeclare{misc}{qiskit}
\bibitem{qiskit}
\bibinfo{author}{H.~\surnamestart Abraham\surnameend}, \bibinfo{author}{I.~Y.
  \surnamestart Akhalwaya\surnameend}, \bibinfo{author}{G.~\surnamestart
  Aleksandrowicz\surnameend}, \bibinfo{author}{T.~\surnamestart
  Alexander\surnameend}, \bibinfo{author}{G.~\surnamestart
  Alexandrowics\surnameend}, \bibinfo{author}{E.~\surnamestart
  Arbel\surnameend}, \bibinfo{author}{A.~\surnamestart Asfaw\surnameend},
  \bibinfo{author}{C.~\surnamestart Azaustre\surnameend},
  \bibinfo{author}{\surnamestart AzizNgoueya\surnameend},
  \bibinfo{author}{P.~\surnamestart Barkoutsos\surnameend},
  \bibinfo{author}{G.~\surnamestart Barron\surnameend},
  \bibinfo{author}{L.~\surnamestart Bello\surnameend},
  \bibinfo{author}{Y.~\surnamestart Ben-Haim\surnameend},
  \bibinfo{author}{D.~\surnamestart Bevenius\surnameend},
  \bibinfo{author}{L.~S. \surnamestart Bishop\surnameend},
  \bibinfo{author}{S.~\surnamestart Bosch\surnameend},
  \bibinfo{author}{D.~\surnamestart Bucher\surnameend},
  \bibinfo{author}{\surnamestart CZ\surnameend},
  \bibinfo{author}{F.~\surnamestart Cabrera\surnameend},
  \bibinfo{author}{P.~\surnamestart Calpin\surnameend},
  \bibinfo{author}{L.~\surnamestart Capelluto\surnameend},
  \bibinfo{author}{J.~\surnamestart Carballo\surnameend},
  \bibinfo{author}{G.~\surnamestart Carrascal\surnameend},
  \bibinfo{author}{A.~\surnamestart Chen\surnameend}, \bibinfo{author}{C.-F.
  \surnamestart Chen\surnameend}, \bibinfo{author}{R.~\surnamestart
  Chen\surnameend}, \bibinfo{author}{J.~M. \surnamestart Chow\surnameend},
  \bibinfo{author}{C.~\surnamestart Claus\surnameend},
  \bibinfo{author}{C.~\surnamestart Clauss\surnameend}, \bibinfo{author}{A.~J.
  \surnamestart Cross\surnameend}, \bibinfo{author}{A.~W. \surnamestart
  Cross\surnameend}, \bibinfo{author}{J.~\surnamestart Cruz-Benito\surnameend},
  \bibinfo{author}{C.~\surnamestart Culver\surnameend}, \bibinfo{author}{A.~D.
  \surnamestart C{\'o}rcoles-Gonzales\surnameend},
  \bibinfo{author}{S.~\surnamestart Dague\surnameend},
  \bibinfo{author}{M.~\surnamestart Dartiailh\surnameend},
  \bibinfo{author}{\surnamestart DavideFrr\surnameend}, \bibinfo{author}{A.~R.
  \surnamestart Davila\surnameend}, \bibinfo{author}{D.~\surnamestart
  Ding\surnameend}, \bibinfo{author}{E.~\surnamestart Drechsler\surnameend},
  \bibinfo{author}{\surnamestart Drew\surnameend},
  \bibinfo{author}{E.~\surnamestart Dumitrescu\surnameend},
  \bibinfo{author}{K.~\surnamestart Dumon\surnameend},
  \bibinfo{author}{I.~\surnamestart Duran\surnameend},
  \bibinfo{author}{P.~\surnamestart Eendebak\surnameend},
  \bibinfo{author}{D.~\surnamestart Egger\surnameend},
  \bibinfo{author}{M.~\surnamestart Everitt\surnameend}, \bibinfo{author}{P.~M.
  \surnamestart Fern{\'a}ndez\surnameend}, \bibinfo{author}{A.~\surnamestart
  Frisch\surnameend}, \bibinfo{author}{A.~\surnamestart Fuhrer\surnameend},
  \bibinfo{author}{M.~\surnamestart GEORGE\surnameend},
  \bibinfo{author}{I.~\surnamestart GOULD\surnameend},
  \bibinfo{author}{J.~\surnamestart Gacon\surnameend},
  \bibinfo{author}{\surnamestart Gadi\surnameend}, \bibinfo{author}{B.~G.
  \surnamestart Gago\surnameend}, \bibinfo{author}{J.~M. \surnamestart
  Gambetta\surnameend}, \bibinfo{author}{L.~\surnamestart Garcia\surnameend},
  \bibinfo{author}{S.~\surnamestart Garion\surnameend},
  \bibinfo{author}{\surnamestart Gawel-Kus\surnameend},
  \bibinfo{author}{J.~\surnamestart Gomez-Mosquera\surnameend},
  \bibinfo{author}{S.~\surnamestart de~la Puente~Gonz{\'a}lez\surnameend},
  \bibinfo{author}{D.~\surnamestart Greenberg\surnameend},
  \bibinfo{author}{W.~\surnamestart Guan\surnameend}, \bibinfo{author}{J.~A.
  \surnamestart Gunnels\surnameend}, \bibinfo{author}{I.~\surnamestart
  Haide\surnameend}, \bibinfo{author}{I.~\surnamestart Hamamura\surnameend},
  \bibinfo{author}{V.~\surnamestart Havlicek\surnameend},
  \bibinfo{author}{J.~\surnamestart Hellmers\surnameend},
  \bibinfo{author}{{\L}.~\surnamestart Herok\surnameend},
  \bibinfo{author}{S.~\surnamestart Hillmich\surnameend},
  \bibinfo{author}{H.~\surnamestart Horii\surnameend},
  \bibinfo{author}{C.~\surnamestart Howington\surnameend},
  \bibinfo{author}{S.~\surnamestart Hu\surnameend},
  \bibinfo{author}{W.~\surnamestart Hu\surnameend},
  \bibinfo{author}{H.~\surnamestart Imai\surnameend},
  \bibinfo{author}{T.~\surnamestart Imamichi\surnameend},
  \bibinfo{author}{R.~\surnamestart Iten\surnameend},
  \bibinfo{author}{T.~\surnamestart Itoko\surnameend},
  \bibinfo{author}{A.~\surnamestart Javadi-Abhari\surnameend},
  \bibinfo{author}{\surnamestart Jessica\surnameend},
  \bibinfo{author}{K.~\surnamestart Johns\surnameend},
  \bibinfo{author}{N.~\surnamestart Kanazawa\surnameend},
  \bibinfo{author}{A.~\surnamestart Karazeev\surnameend},
  \bibinfo{author}{P.~\surnamestart Kassebaum\surnameend},
  \bibinfo{author}{\surnamestart Knabberjoe\surnameend},
  \bibinfo{author}{A.~\surnamestart Kovyrshin\surnameend},
  \bibinfo{author}{V.~\surnamestart Krishnan\surnameend},
  \bibinfo{author}{K.~\surnamestart Krsulich\surnameend},
  \bibinfo{author}{G.~\surnamestart Kus\surnameend},
  \bibinfo{author}{R.~\surnamestart LaRose\surnameend},
  \bibinfo{author}{R.~\surnamestart Lambert\surnameend},
  \bibinfo{author}{J.~\surnamestart Latone\surnameend},
  \bibinfo{author}{S.~\surnamestart Lawrence\surnameend},
  \bibinfo{author}{D.~\surnamestart Liu\surnameend},
  \bibinfo{author}{P.~\surnamestart Liu\surnameend}, \bibinfo{author}{P.~B.~Z.
  \surnamestart Mac\surnameend}, \bibinfo{author}{Y.~\surnamestart
  Maeng\surnameend}, \bibinfo{author}{A.~\surnamestart Malyshev\surnameend},
  \bibinfo{author}{J.~\surnamestart Marecek\surnameend},
  \bibinfo{author}{M.~\surnamestart Marques\surnameend},
  \bibinfo{author}{D.~\surnamestart Mathews\surnameend},
  \bibinfo{author}{A.~\surnamestart Matsuo\surnameend}, \bibinfo{author}{D.~T.
  \surnamestart McClure\surnameend}, \bibinfo{author}{C.~\surnamestart
  McGarry\surnameend}, \bibinfo{author}{D.~\surnamestart McKay\surnameend},
  \bibinfo{author}{S.~\surnamestart Meesala\surnameend},
  \bibinfo{author}{A.~\surnamestart Mezzacapo\surnameend},
  \bibinfo{author}{R.~\surnamestart Midha\surnameend},
  \bibinfo{author}{Z.~\surnamestart Minev\surnameend}, \bibinfo{author}{M.~D.
  \surnamestart Mooring\surnameend}, \bibinfo{author}{R.~\surnamestart
  Morales\surnameend}, \bibinfo{author}{N.~\surnamestart Moran\surnameend},
  \bibinfo{author}{P.~\surnamestart Murali\surnameend},
  \bibinfo{author}{J.~\surnamestart M{\"u}ggenburg\surnameend},
  \bibinfo{author}{D.~\surnamestart Nadlinger\surnameend},
  \bibinfo{author}{G.~\surnamestart Nannicini\surnameend},
  \bibinfo{author}{P.~\surnamestart Nation\surnameend},
  \bibinfo{author}{Y.~\surnamestart Naveh\surnameend},
  \bibinfo{author}{\surnamestart Nick-Singstock\surnameend},
  \bibinfo{author}{P.~\surnamestart Niroula\surnameend},
  \bibinfo{author}{H.~\surnamestart Norlen\surnameend}, \bibinfo{author}{L.~J.
  \surnamestart O'Riordan\surnameend}, \bibinfo{author}{O.~\surnamestart
  Ogunbayo\surnameend}, \bibinfo{author}{P.~\surnamestart
  Ollitrault\surnameend}, \bibinfo{author}{S.~\surnamestart Oud\surnameend},
  \bibinfo{author}{D.~\surnamestart Padilha\surnameend},
  \bibinfo{author}{H.~\surnamestart Paik\surnameend},
  \bibinfo{author}{S.~\surnamestart Perriello\surnameend},
  \bibinfo{author}{A.~\surnamestart Phan\surnameend},
  \bibinfo{author}{M.~\surnamestart Pistoia\surnameend},
  \bibinfo{author}{A.~\surnamestart Pozas-iKerstjens\surnameend},
  \bibinfo{author}{V.~\surnamestart Prutyanov\surnameend},
  \bibinfo{author}{D.~\surnamestart Puzzuoli\surnameend},
  \bibinfo{author}{J.~\surnamestart P{\'e}rez\surnameend},
  \bibinfo{author}{\surnamestart Quintiii\surnameend},
  \bibinfo{author}{R.~\surnamestart Raymond\surnameend},
  \bibinfo{author}{R.~M.-C. \surnamestart Redondo\surnameend},
  \bibinfo{author}{M.~\surnamestart Reuter\surnameend}, \bibinfo{author}{D.~M.
  \surnamestart Rodr{\'\i}guez\surnameend}, \bibinfo{author}{M.~\surnamestart
  Ryu\surnameend}, \bibinfo{author}{T.~\surnamestart SAPV\surnameend},
  \bibinfo{author}{\surnamestart SamFerracin\surnameend},
  \bibinfo{author}{M.~\surnamestart Sandberg\surnameend},
  \bibinfo{author}{N.~\surnamestart Sathaye\surnameend},
  \bibinfo{author}{B.~\surnamestart Schmitt\surnameend},
  \bibinfo{author}{C.~\surnamestart Schnabel\surnameend},
  \bibinfo{author}{T.~L. \surnamestart Scholten\surnameend},
  \bibinfo{author}{E.~\surnamestart Schoute\surnameend}, \bibinfo{author}{I.~F.
  \surnamestart Sertage\surnameend}, \bibinfo{author}{N.~\surnamestart
  Shammah\surnameend}, \bibinfo{author}{Y.~\surnamestart Shi\surnameend},
  \bibinfo{author}{A.~\surnamestart Silva\surnameend},
  \bibinfo{author}{Y.~\surnamestart Siraichi\surnameend},
  \bibinfo{author}{I.~\surnamestart Sitdikov\surnameend},
  \bibinfo{author}{S.~\surnamestart Sivarajah\surnameend},
  \bibinfo{author}{J.~A. \surnamestart Smolin\surnameend},
  \bibinfo{author}{M.~\surnamestart Soeken\surnameend},
  \bibinfo{author}{\surnamestart SooluThomas\surnameend},
  \bibinfo{author}{D.~\surnamestart Steenken\surnameend},
  \bibinfo{author}{M.~\surnamestart Stypulkoski\surnameend},
  \bibinfo{author}{H.~\surnamestart Takahashi\surnameend},
  \bibinfo{author}{C.~\surnamestart Taylor\surnameend},
  \bibinfo{author}{P.~\surnamestart Taylour\surnameend},
  \bibinfo{author}{S.~\surnamestart Thomas\surnameend},
  \bibinfo{author}{M.~\surnamestart Tillet\surnameend},
  \bibinfo{author}{M.~\surnamestart Tod\surnameend},
  \bibinfo{author}{E.~\surnamestart de~la Torre\surnameend},
  \bibinfo{author}{K.~\surnamestart Trabing\surnameend},
  \bibinfo{author}{M.~\surnamestart Treinish\surnameend},
  \bibinfo{author}{\surnamestart TrishaPe\surnameend},
  \bibinfo{author}{W.~\surnamestart Turner\surnameend},
  \bibinfo{author}{Y.~\surnamestart Vaknin\surnameend}, \bibinfo{author}{C.~R.
  \surnamestart Valcarce\surnameend}, \bibinfo{author}{F.~\surnamestart
  Varchon\surnameend}, \bibinfo{author}{D.~\surnamestart Vogt-Lee\surnameend},
  \bibinfo{author}{C.~\surnamestart Vuillot\surnameend},
  \bibinfo{author}{J.~\surnamestart Weaver\surnameend},
  \bibinfo{author}{R.~\surnamestart Wieczorek\surnameend},
  \bibinfo{author}{J.~A. \surnamestart Wildstrom\surnameend},
  \bibinfo{author}{R.~\surnamestart Wille\surnameend},
  \bibinfo{author}{E.~\surnamestart Winston\surnameend}, \bibinfo{author}{J.~J.
  \surnamestart Woehr\surnameend}, \bibinfo{author}{S.~\surnamestart
  Woerner\surnameend}, \bibinfo{author}{R.~\surnamestart Woo\surnameend},
  \bibinfo{author}{C.~J. \surnamestart Wood\surnameend},
  \bibinfo{author}{R.~\surnamestart Wood\surnameend},
  \bibinfo{author}{S.~\surnamestart Wood\surnameend},
  \bibinfo{author}{J.~\surnamestart Wootton\surnameend},
  \bibinfo{author}{D.~\surnamestart Yeralin\surnameend},
  \bibinfo{author}{J.~\surnamestart Yu\surnameend},
  \bibinfo{author}{C.~\surnamestart Zachow\surnameend},
  \bibinfo{author}{L.~\surnamestart Zdanski\surnameend},
  \bibinfo{author}{\surnamestart Zoufalc\surnameend},
  \bibinfo{author}{\surnamestart anedumla\surnameend},
  \bibinfo{author}{\surnamestart azulehner\surnameend},
  \bibinfo{author}{\surnamestart bcamorrison\surnameend},
  \bibinfo{author}{\surnamestart brandhsn\surnameend},
  \bibinfo{author}{\surnamestart dennis-liu 1\surnameend},
  \bibinfo{author}{\surnamestart dime10\surnameend},
  \bibinfo{author}{\surnamestart drholmie\surnameend},
  \bibinfo{author}{\surnamestart elfrocampeador\surnameend},
  \bibinfo{author}{\surnamestart faisaldebouni\surnameend},
  \bibinfo{author}{\surnamestart fanizzamarco\surnameend},
  \bibinfo{author}{\surnamestart gruu\surnameend},
  \bibinfo{author}{\surnamestart kanejess\surnameend},
  \bibinfo{author}{\surnamestart klinvill\surnameend},
  \bibinfo{author}{\surnamestart kurarrr\surnameend},
  \bibinfo{author}{\surnamestart lerongil\surnameend},
  \bibinfo{author}{\surnamestart ma5x\surnameend},
  \bibinfo{author}{\surnamestart merav aharoni\surnameend},
  \bibinfo{author}{\surnamestart mrossinek\surnameend},
  \bibinfo{author}{\surnamestart ordmoj\surnameend},
  \bibinfo{author}{\surnamestart strickroman\surnameend},
  \bibinfo{author}{\surnamestart tigerjack\surnameend},
  \bibinfo{author}{\surnamestart toural\surnameend},
  \bibinfo{author}{\surnamestart yang.luh\surnameend} \&
  \bibinfo{author}{\surnamestart yotamvakninibm\surnameend}
  (\bibinfo{year}{2019}): \emph{\bibinfo{title}{Qiskit: An Open-source
  Framework for Quantum Computing}}, \doi{10.5281/zenodo.2562110}.

\bibitemdeclare{misc}{basilewitsch2020}
\bibitem{basilewitsch2020}
\bibinfo{author}{D.~\surnamestart Basilewitsch\surnameend},
  \bibinfo{author}{J.~\surnamestart Fischer\surnameend}, \bibinfo{author}{D.~M.
  \surnamestart Reich\surnameend}, \bibinfo{author}{D.~\surnamestart
  Sugny\surnameend} \& \bibinfo{author}{C.~P. \surnamestart Koch\surnameend}
  (\bibinfo{year}{2020}): \emph{\bibinfo{title}{Fundamental Bounds on Qubit
  Reset}}.
\newblock \urlprefix\url{https://arxiv.org/abs/2001.09107}.

\bibitemdeclare{techreport}{NIST}
\bibitem{NIST}
\bibinfo{author}{L.~\surnamestart Bassham\surnameend},
  \bibinfo{author}{A.~\surnamestart Rukhin\surnameend},
  \bibinfo{author}{J.~\surnamestart Soto\surnameend},
  \bibinfo{author}{J.~\surnamestart Nechvatal\surnameend},
  \bibinfo{author}{M.~\surnamestart Smid\surnameend},
  \bibinfo{author}{E.~\surnamestart Barker\surnameend},
  \bibinfo{author}{S.~\surnamestart Leigh\surnameend},
  \bibinfo{author}{M.~\surnamestart Levenson\surnameend},
  \bibinfo{author}{M.~\surnamestart Vangel\surnameend},
  \bibinfo{author}{D.~\surnamestart Banks\surnameend},
  \bibinfo{author}{N.~\surnamestart Heckert\surnameend} \&
  \bibinfo{author}{J.~\surnamestart Dray\surnameend} (\bibinfo{year}{2010}):
  \emph{\bibinfo{title}{A Statistical Test Suite for Random and Pseudorandom
  Number Generators for Cryptographic Applications}}.
\newblock \bibinfo{type}{NIST SP 800-22 Rev. 1a},
  \bibinfo{institution}{National Institute of Standards and Technology},
  \doi{10.6028/NIST.SP.800-22r1a}.

\bibitemdeclare{article}{Burnett2019}
\bibitem{Burnett2019}
\bibinfo{author}{J.~J. \surnamestart Burnett\surnameend},
  \bibinfo{author}{A.~\surnamestart Bengtsson\surnameend},
  \bibinfo{author}{M.~\surnamestart Scigliuzzo\surnameend},
  \bibinfo{author}{D.~\surnamestart Niepce\surnameend},
  \bibinfo{author}{M.~\surnamestart Kudra\surnameend},
  \bibinfo{author}{P.~\surnamestart Delsing\surnameend} \&
  \bibinfo{author}{J.~\surnamestart Bylander\surnameend}
  (\bibinfo{year}{2019}): \emph{\bibinfo{title}{Decoherence benchmarking of
  superconducting qubits}}.
\newblock {\sl \bibinfo{journal}{npj Quantum Information}} \bibinfo{volume}{5},
  p.~\bibinfo{pages}{54}, \doi{10.1038/s41534-019-0168-5}.

\bibitemdeclare{article}{Dewes2012}
\bibitem{Dewes2012}
\bibinfo{author}{A.~\surnamestart Dewes\surnameend},
  \bibinfo{author}{R.~\surnamestart Lauro\surnameend}, \bibinfo{author}{F.~R.
  \surnamestart Ong\surnameend}, \bibinfo{author}{V.~\surnamestart
  Schmitt\surnameend}, \bibinfo{author}{P.~\surnamestart Milman\surnameend},
  \bibinfo{author}{P.~\surnamestart Bertet\surnameend},
  \bibinfo{author}{D.~\surnamestart Vion\surnameend} \&
  \bibinfo{author}{D.~\surnamestart Esteve\surnameend} (\bibinfo{year}{2012}):
  \emph{\bibinfo{title}{Quantum speeding-up of computation demonstrated in a
  superconducting two-qubit processor}}.
\newblock {\sl \bibinfo{journal}{Phys. Rev. B}} \bibinfo{volume}{85}, p.
  \bibinfo{pages}{140503}, \doi{10.1103/PhysRevB.85.140503}.

\bibitemdeclare{article}{Egger2018}
\bibitem{Egger2018}
\bibinfo{author}{D.~J. \surnamestart Egger\surnameend},
  \bibinfo{author}{M.~\surnamestart Werninghaus\surnameend},
  \bibinfo{author}{M.~\surnamestart Ganzhorn\surnameend},
  \bibinfo{author}{G.~\surnamestart Salis\surnameend},
  \bibinfo{author}{A.~\surnamestart Fuhrer\surnameend},
  \bibinfo{author}{P.~\surnamestart M\"uller\surnameend} \&
  \bibinfo{author}{S.~\surnamestart Filipp\surnameend} (\bibinfo{year}{2018}):
  \emph{\bibinfo{title}{Pulsed Reset Protocol for Fixed-Frequency
  Superconducting Qubits}}.
\newblock {\sl \bibinfo{journal}{Physical Review Applied}}
  \bibinfo{volume}{10}, p. \bibinfo{pages}{044030},
  \doi{10.1103/PhysRevApplied.10.044030}.

\bibitemdeclare{misc}{Eisert2019}
\bibitem{Eisert2019}
\bibinfo{author}{J.~\surnamestart Eisert\surnameend},
  \bibinfo{author}{D.~\surnamestart Hangleiter\surnameend},
  \bibinfo{author}{N.~\surnamestart Walk\surnameend},
  \bibinfo{author}{I.~\surnamestart Roth\surnameend},
  \bibinfo{author}{D.~\surnamestart Markham\surnameend},
  \bibinfo{author}{R.~\surnamestart Parekh\surnameend},
  \bibinfo{author}{U.~\surnamestart Chabaud\surnameend} \&
  \bibinfo{author}{E.~\surnamestart Kashefi\surnameend} (\bibinfo{year}{2019}):
  \emph{\bibinfo{title}{Quantum certification and benchmarking}}.
\newblock \urlprefix\url{https://arxiv.org/abs/1910.06343}.

\bibitemdeclare{article}{Herrero-Collantes2017}
\bibitem{Herrero-Collantes2017}
\bibinfo{author}{M.~\surnamestart Herrero-Collantes\surnameend} \&
  \bibinfo{author}{J.~C. \surnamestart Garcia-Escartin\surnameend}
  (\bibinfo{year}{2017}): \emph{\bibinfo{title}{Quantum random number
  generators}}.
\newblock {\sl \bibinfo{journal}{Reviews of Modern Physics}}
  \bibinfo{volume}{89}, p. \bibinfo{pages}{015004},
  \doi{10.1103/RevModPhys.89.015004}.

\bibitemdeclare{article}{Hosoya2011}
\bibitem{Hosoya2011}
\bibinfo{author}{A.~\surnamestart Hosoya\surnameend},
  \bibinfo{author}{K.~\surnamestart Maruyama\surnameend} \&
  \bibinfo{author}{Y.~\surnamestart Shikano\surnameend} (\bibinfo{year}{2011}):
  \emph{\bibinfo{title}{Maxwell's demon and data compression}}.
\newblock {\sl \bibinfo{journal}{Physical Review E}} \bibinfo{volume}{84}, p.
  \bibinfo{pages}{061117}, \doi{10.1103/PhysRevE.84.061117}.

\bibitemdeclare{article}{Kak1999}
\bibitem{Kak1999}
\bibinfo{author}{S.~\surnamestart Kak\surnameend} (\bibinfo{year}{1999}):
  \emph{\bibinfo{title}{The Initialization Problem in Quantum Computing}}.
\newblock {\sl \bibinfo{journal}{Foundations of Physics}} \bibinfo{volume}{29},
  p. \bibinfo{pages}{267–279}, \doi{10.1023/A:1018877706849}.

\bibitemdeclare{article}{Landauer}
\bibitem{Landauer}
\bibinfo{author}{R.~\surnamestart Landauer\surnameend} (\bibinfo{year}{1961}):
  \emph{\bibinfo{title}{Irreversibility and heat generation in the computing
  process}}.
\newblock {\sl \bibinfo{journal}{IBM Journal of Research and Development}}
  \bibinfo{volume}{5}, pp. \bibinfo{pages}{183--191}, \doi{10.1147/rd.53.0183}.

\bibitemdeclare{article}{testu01}
\bibitem{testu01}
\bibinfo{author}{P.~\surnamestart L'Ecuyer\surnameend} \&
  \bibinfo{author}{R.~\surnamestart Simard\surnameend} (\bibinfo{year}{2007}):
  \emph{\bibinfo{title}{TestU01: A C Library for Empirical Testing of Random
  Number Generators}}.
\newblock {\sl \bibinfo{journal}{ACM Transactions on Mathematical Software
  (TOMS)}} \bibinfo{volume}{33}(\bibinfo{number}{4}), p.~\bibinfo{pages}{22},
  \doi{10.1145/1268776.1268777}.

\bibitemdeclare{article}{Ma2016}
\bibitem{Ma2016}
\bibinfo{author}{X.~\surnamestart Ma\surnameend},
  \bibinfo{author}{X.~\surnamestart Yuan\surnameend},
  \bibinfo{author}{Z.~\surnamestart Cao\surnameend},
  \bibinfo{author}{B.~\surnamestart Qi\surnameend} \&
  \bibinfo{author}{Z.~\surnamestart Zhang\surnameend} (\bibinfo{year}{2016}):
  \emph{\bibinfo{title}{Quantum random number generation}}.
\newblock {\sl \bibinfo{journal}{npj Quantum Information}} \bibinfo{volume}{2},
  p. \bibinfo{pages}{16021}, \doi{10.1038/npjqi.2016.21}.

\bibitemdeclare{article}{Magnard2018}
\bibitem{Magnard2018}
\bibinfo{author}{P.~\surnamestart Magnard\surnameend},
  \bibinfo{author}{P.~\surnamestart Kurpiers\surnameend},
  \bibinfo{author}{B.~\surnamestart Royer\surnameend},
  \bibinfo{author}{T.~\surnamestart Walter\surnameend}, \bibinfo{author}{J.-C.
  \surnamestart Besse\surnameend}, \bibinfo{author}{S.~\surnamestart
  Gasparinetti\surnameend}, \bibinfo{author}{M.~\surnamestart
  Pechal\surnameend}, \bibinfo{author}{J.~\surnamestart Heinsoo\surnameend},
  \bibinfo{author}{S.~\surnamestart Storz\surnameend},
  \bibinfo{author}{A.~\surnamestart Blais\surnameend} \&
  \bibinfo{author}{A.~\surnamestart Wallraff\surnameend}
  (\bibinfo{year}{2018}): \emph{\bibinfo{title}{Fast and Unconditional
  All-Microwave Reset of a Superconducting Qubit}}.
\newblock {\sl \bibinfo{journal}{Physical Review Letters}}
  \bibinfo{volume}{121}, p. \bibinfo{pages}{060502},
  \doi{10.1103/PhysRevLett.121.060502}.

\bibitemdeclare{article}{Petta2005}
\bibitem{Petta2005}
\bibinfo{author}{J.~R. \surnamestart Petta\surnameend}, \bibinfo{author}{A.~C.
  \surnamestart Johnson\surnameend}, \bibinfo{author}{J.~M. \surnamestart
  Taylor\surnameend}, \bibinfo{author}{E.~A. \surnamestart Laird\surnameend},
  \bibinfo{author}{A.~\surnamestart Yacoby\surnameend}, \bibinfo{author}{M.~D.
  \surnamestart Lukin\surnameend}, \bibinfo{author}{C.~M. \surnamestart
  Marcus\surnameend}, \bibinfo{author}{M.~P. \surnamestart Hanson\surnameend}
  \& \bibinfo{author}{A.~C. \surnamestart Gossard\surnameend}
  (\bibinfo{year}{2005}): \emph{\bibinfo{title}{Coherent Manipulation of
  Coupled Electron Spins in Semiconductor Quantum Dots}}.
\newblock {\sl \bibinfo{journal}{Science}} \bibinfo{volume}{309}, pp.
  \bibinfo{pages}{2180--2184}, \doi{10.1126/science.1116955}.

\bibitemdeclare{article}{Riste2012}
\bibitem{Riste2012}
\bibinfo{author}{D.~\surnamestart Rist\`e\surnameend}, \bibinfo{author}{J.~G.
  \surnamestart van Leeuwen\surnameend}, \bibinfo{author}{H.-S. \surnamestart
  Ku\surnameend}, \bibinfo{author}{K.~W. \surnamestart Lehnert\surnameend} \&
  \bibinfo{author}{L.~\surnamestart DiCarlo\surnameend} (\bibinfo{year}{2012}):
  \emph{\bibinfo{title}{Initialization by Measurement of a Superconducting
  Quantum Bit Circuit}}.
\newblock {\sl \bibinfo{journal}{Physical Review Letters}}
  \bibinfo{volume}{109}, p. \bibinfo{pages}{050507},
  \doi{10.1103/PhysRevLett.109.050507}.

\bibitemdeclare{inproceedings}{tamura2019}
\bibitem{tamura2019}
\bibinfo{author}{K.~\surnamestart Tamura\surnameend} \&
  \bibinfo{author}{Y.~\surnamestart Shikano\surnameend} (\bibinfo{year}{2019}):
  \emph{\bibinfo{title}{Quantum Random Number Generation with the
  Superconducting Quantum Computer IBM 20Q Tokyo}}.
\newblock In \bibinfo{editor}{M.~\surnamestart Hirvensalo\surnameend} \&
  \bibinfo{editor}{A.~\surnamestart Yakaryılmaz\surnameend}, editors: {\sl
  \bibinfo{booktitle}{Proceedings of Workshop on Quantum Computing and Quantum
  Information}}, {\sl \bibinfo{series}{TUCS Lecture
  Notes}}~\bibinfo{volume}{30}, pp. \bibinfo{pages}{13--25}.
\newblock \urlprefix\url{http://urn.fi/URN:ISBN:978-952-12-3840-6}.
\newblock \bibinfo{note}{Cryptology ePrint Archive, Report 2020/078
  \url{https://eprint.iacr.org/2020/078}}.

\bibitemdeclare{inproceedings}{tamura2020}
\bibitem{tamura2020}
\bibinfo{author}{K.~\surnamestart Tamura\surnameend} \&
  \bibinfo{author}{Y.~\surnamestart Shikano\surnameend} (\bibinfo{year}{2020}):
  \emph{\bibinfo{title}{Quantum Random Numbers generated by the Cloud
  Superconducting Quantum Computer}}.
\newblock In \bibinfo{editor}{T.~\surnamestart Takagi\surnameend},
  \bibinfo{editor}{M.~\surnamestart Wakayama\surnameend},
  \bibinfo{editor}{K.~\surnamestart Tanaka\surnameend},
  \bibinfo{editor}{N.~\surnamestart Kunihiro\surnameend},
  \bibinfo{editor}{K.~\surnamestart Kimoto\surnameend} \&
  \bibinfo{editor}{Y.~\surnamestart Ikematsu\surnameend}, editors: {\sl
  \bibinfo{booktitle}{International Symposium on Mathematics, Quantum Theory,
  and Cryptography: Proceedings of MQC 2019}}, \bibinfo{publisher}{Springer
  Nature}.
\newblock \urlprefix\url{https://arxiv.org/abs/1906.04410}.
\newblock \bibinfo{note}{To be published, arXiv:1906.04410}.

\bibitemdeclare{article}{Tuorila2017}
\bibitem{Tuorila2017}
\bibinfo{author}{J.~\surnamestart Tuorila\surnameend},
  \bibinfo{author}{M.~\surnamestart Partanen\surnameend},
  \bibinfo{author}{T.~\surnamestart Ala-Nissila\surnameend} \&
  \bibinfo{author}{M.~\surnamestart M\"ott\"onen\surnameend}
  (\bibinfo{year}{2017}): \emph{\bibinfo{title}{Efficient protocol for qubit
  initialization with a tunable environment}}.
\newblock {\sl \bibinfo{journal}{npj Quantum Information}} \bibinfo{volume}{3},
  p.~\bibinfo{pages}{27}, \doi{10.1038/s41534-017-0027-1}.

\end{thebibliography}
\end{document}